\documentclass[prd,showpacs,showkeys,nofootinbib,twocolumn]{revtex4-1}

\usepackage{graphicx}
\usepackage{amsmath,amsfonts,amssymb,amsxtra,latexsym}

\usepackage{hyperref}
\hypersetup{
    colorlinks=true,
    linkcolor=red,
    citecolor=blue,      
    urlcolor=blue,
}

\usepackage[utf8]{inputenc}

\begin{document}

\title{Bonnor--Vaidya Charged Point Mass in an External Maxwell Field}

\author{Peter A. Hogan}
\email{peter.hogan@ucd.ie}
\affiliation{School of Physics, University College Dublin, Belfield, Dublin 4, Ireland} 

\author{Dirk Puetzfeld}
\email{dirk.puetzfeld@zarm.uni-bremen.de}
\homepage{http://puetzfeld.org}
\affiliation{University of Bremen, Center of Applied Space Technology and Microgravity (ZARM), 28359 Bremen, Germany} 

\date{ \today}

\begin{abstract}
By introducing external Maxwell and gravitational fields we modify the Bonnor--Vaidya field of an arbitrarily accelerating charged mass moving rectilinearly in order to satisfy the vacuum Einstein--Maxwell field equations approximately, assuming the charge $e$ and the mass $m$ are small of first order.
\end{abstract}


\maketitle


\section{Introduction}\label{sec_introduction}

\begin{figure}
\begin{center}
\includegraphics[height=8cm]{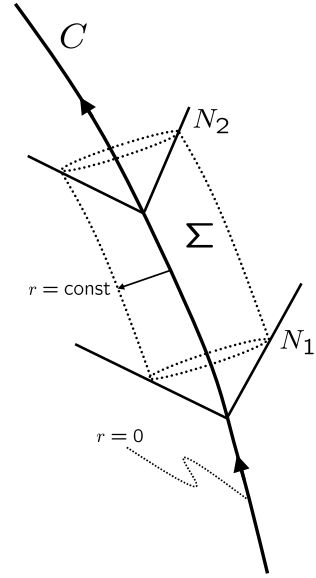}
\end{center}
\caption{\label{fig_1} The world line $C(r=0)$ in the background space--time. $\Sigma$ is a world tube $r={\rm const}>0$ bounded by the future null cones $N_1(u=u_1)$ and $N_2(u=u_2)$ with $u_2>u_1$ constants.}
\end{figure}

The solution of Einstein's field equations describing a space--time model of an arbitrarily accelerating point mass found by Kinnersley \cite{Kinnersley:1969} has been described as a photon rocket by Bonnor \cite{Bonnor:1994}. Kinnersley's metric tensor is of Kerr--Schild form \cite{Kerr:Schild:1965} and thus possesses a background Minkowskian space--time obtained by putting the mass of the source equal to zero. In this background the source of the field is an arbitrary timelike world line. From this point of view the 4--momentum radiated during a finite interval of proper time is given exactly by the change in the particle 4--momentum during this interval (a ``rocket effect" described in detail by Bonnor \cite{Bonnor:1994}). In the background Minkowskian space--time picture this radiated 4--momentum is a flux of 4--momentum across a timelike world tube surrounding the particle world line and bounded by two future null cones with vertices on the world line separated by a finite interval of proper time (see Fig.\ \ref{fig_1}). It follows that the particle is self accelerated by photon emission and is therefore referred to as a photon rocket. An extension of the Kinnersley rocket to include charge has been given by Bonnor and Vaidya \cite{Bonnor:Vaidya:1972} (we assume for simplicity that the mass and charge are both constant). This Bonnor--Vaidya particle is also self accelerating via photon emission (with no loss of charge). In the present paper we consider a Bonnor--Vaidya particle performing rectilinear motion but driven by a suitable external Maxwell field. The Einstein--Maxwell field equations are solved approximately assuming that the mass $m$ and the charge $e$ of the particle are both constant and small of first order. The field equations are solved up to the second order of approximation which involves working with an error of third order in these small quantities. We demonstrate that the rocket effect can be removed, at least to the order of approximation that we are working, and instead the particle is driven by the external field.

To make clear the background to the present study we point out here that Kinnersley's field \cite{Kinnersley:1969} of an arbitrarily accelerating point mass, in the special case of rectilinear motion, is described by the line element
\begin{eqnarray}
ds^2&=&-r^2\{(d\theta+A(u)\,\sin\theta\,du)^2+\sin^2\theta\,d\phi^2\}+2\,du\,dr\nonumber\\
&&+\left (1-2\,A(u)\,r\,\cos\theta-\frac{2\,m}{r}\right )\,du^2\ .\label{1}
\end{eqnarray}
The constant $m$ is the mass of the particle and $A(u)$ is the arbitrary acceleration. The corresponding Ricci tensor components $R_{i'j'}$ in the coordinates $x^{i'}=(\theta, \phi, r, u)$ with $i'=1', 2', 3', 4'$ (we reserve unprimed indices for labelling rectangular Cartesian coordinates and time later) take the lightlike dust or Vaidya form
\begin{equation}\label{2}
R_{i'j'}=-\frac{6\,m\,A(u)\,\cos\theta}{r^2}n_{i'}\,n_{j'}=-8\,\pi\,T_{i'j'}\ ,
\end{equation}
with $n_{i'}\,dx^{i'}=du$ and $n^{i'}\partial/\partial x^{i'}=\partial/\partial r$ is a null vector field. Charged generalizations have been given by Bonnor and Vaidya \cite{Bonnor:Vaidya:1972}. The simplest example is
\begin{eqnarray}
ds^2&=&g_{i'j'}\,dx^{i'}\,dx^{j'}\ ,\nonumber\\
&=&-r^2\{(d\theta+A(u)\,\sin\theta\,du)^2+\sin^2\theta\,d\phi^2\}+2\,du\,dr\nonumber\\
&&+\left (1-2\,A(u)\,r\,\cos\theta-\frac{2\,m}{r}+\frac{e^2}{r^2}\right )\,du^2\ ,\label{3}
\end{eqnarray}
with the Maxwell 2--form
\begin{equation}\label{4}
\frac{1}{2}F_{i'j'}\,dx^{i'}\wedge dx^{j'}=\frac{e}{r^2}\,du\wedge dr\ ,\end{equation}
and $e={\rm constant}$ is the charge on the accelerating particle. When $A(u)=0$ this coincides with the Reissner--Nordstr\"om solution of the vacuum Einstein--Maxwell equations. With the electromagnetic energy--momentum tensor given by 
\begin{equation}\label{5}
E_{i'j'}=F_{i'k'}\,F_{j'}{}^{k'}-\frac{1}{4}\,g_{i'j'}\,F_{l'k'}\,F^{l'k'}\ ,
\end{equation}
the Einstein--Maxwell field equations for (\ref{3}) and (\ref{4}) read
\begin{eqnarray}
&&R_{i'j'}-2\,E_{i'j'}= -8\,\pi\,T_{i'j'}\ \nonumber \\
&&=\left (-\frac{6\,m\,A(u)\,\cos\theta}{r^2}+\frac{4\,e^2A(u)\,\cos\theta}{r^3}\right )\,n_{i'}\,n_{j'}\ ,\label{6}
\end{eqnarray}  
and
\begin{equation}\label{7}
F^{i'j'}{}_{;j'}=J^{i'}=-\frac{2\,e\,A(u)\,\cos\theta}{r^2}\,n^{i'}\ ,
\end{equation}
with the semicolon denoting covariant differentiation with the respect to the Riemannian connection calculated with the metric tensor given via (\ref{3}). In (\ref{2}), (\ref{6}) and (\ref{7}) the resulting matter distribution is described by an energy--momentum--stress tensor with components $T_{i'j'}$ and a 4--current $J^{i'}$.

The organization of the paper is as follows: In section \ref{sec_external_fields} the axially symmetric background space--time is constructed in the neighborhood of a timelike world line which is the history of a particle performing rectilinear motion with arbitrary acceleration. The Einstein--Maxwell field equations are solved in the neighborhood of this world line with a Maxwell field which specializes to a pure electric field on the world line. In section \ref{sec_charged_mass_particle} the charged particle is introduced as a perturbation of the background space--time which is singular on the world line but is otherwise a well behaved perturbation. This latter requirement places an important constraint on the acceleration of the particle while solving approximately the perturbed Einstein--Maxwell field equations. The acceleration of the particle is no longer arbitrary but is driven by the external electric field. Since the perturbed field equations are solved approximately there is a residual matter distribution present which is described by a residual energy--momentum--stress tensor and a residual 4--current. These are examined in section \ref{sec_res_matter_dist} and interpreted physically in terms of the flow of 4--momentum and charge away from the particle. The paper ends with a brief comparison of our model with the Bonnor--Vaidya model in section \ref{sec_discussion}.  

\section{External Fields}\label{sec_external_fields}

\begin{figure}
    \begin{center}
    \includegraphics[height=8cm]{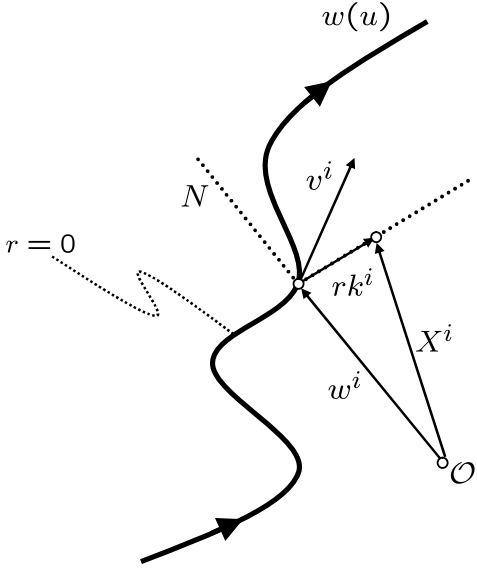}
    \end{center}
    \caption{\label{fig_2} Coordinates of an event $X^i$ in the neighborhood of the world line $w^i(u)$. Here $k^i$ denotes a future pointing null vector field, the parameter value $r=0$ corresponds to events on $w^i(u)$.}
\end{figure}

The external gravitational and electromagnetic fields will be modelled by a space--time and a Maxwell field which will be solutions of the vacuum Einstein--Maxwell field equations. The accelerating charged particle will have a timelike world line in this space--time. The particle will be introduced as a perturbation of this ``background" space--time which is singular on this world line but whose electromagnetic and gravitational fields are otherwise free of singularities. The Lorentzian character of the background space--time means that in the neighborhood of the particle world line the space--time is Minkowskian. This means that if $r$ is a distance from the world line, and if the world line corresponds to $r=0$, then the metric tensor of the background, in rectangular Cartesian coordinates and time $X^i=(X, Y, Z, T)$, satisfies
\begin{equation}\label{8}
g_{ij}=\eta_{ij}+O(r^2)\ ,
\end{equation}
for small values of $r$ with $\eta_{ij}={\rm diag}(-1, -1, -1, +1)$. Since we are interested in a particle performing rectilinear motion we will take it to be moving on the $Z$--axis and thus have a world line in the $(Z, T)$--plane in the Minkowskian space--time neighborhood of its world line. The world line $r=0$ will be given parametrically by $X^i=w^i(u)$ with $w^i(u)=(0, 0, w^3(u), w^4(u))$. The unit timelike tangent to the world line is $v^i(u)=dw^i(u)/du$ with $\eta_{ij}\,v^i\,v^j=-(v^3)^2+(v^4)^2=+1$ so that the parameter $u$, for which $-\infty<u<+\infty$, is proper time or arc length along the world line. The 4--acceleration of the particle with world line $r=0$ is $a^i(u)=dv^i/du$ and, since $v_i\,v^i=\eta_{ij}\,v^i\,v^j=+1$, we have $v_i\,a^i=0$ indicating that the 4--acceleration is spacelike. Defining $A(u)=\{(a^3)^2-(a^4)^2\}^{1/2}$ which we shall refer to as the acceleration of the particle performing rectilinear motion we note that $a^3=A\,v^4$ and $a^4=A\,v^3$. The position 4--vector of an event in the neighborhood of the world line, i.e.\ for small positive values of $r\geq 0$, can be written (see, for example \cite{Newman:Unti:1962})
\begin{equation}\label{9}
X^i=w^i(u)+r\,k^i\ ,
\end{equation}
where $k^i$ is a future pointing null vector field, cf.\ Fig.\ \ref{fig_2}, parametrized by the polar angles $\theta, \phi$ (for which $0\leq\theta\leq\pi$ and $0\leq\phi<2\,\pi$) and normalized by the condition $v_i\,k^i=+1$ so that we may satisfy these requirements by writing (cf.\ \cite{Hogan:Puetzfeld:2020})
\begin{eqnarray}\label{10}
k^i&=&\Big(-\sin\theta\,\cos\phi, -\sin\theta\,\sin\phi, \nonumber \\ 
&& v^3-v^4\,\cos\theta, v^4-v^3\,\cos\theta\Big)\ .
\end{eqnarray}

The following Minkowskian scalar products are useful:
\begin{equation}\label{11}
\frac{\partial k^i}{\partial\theta}\,\frac{\partial k_i}{\partial\theta}=-1\ ,\frac{\partial k^i}{\partial\phi}\,\frac{\partial k_i}{\partial\phi}=-\sin^2\theta\ , \frac{\partial k^i}{\partial\theta}\,\frac{\partial k_i}{\partial\phi}=0\ ,
\end{equation}
and
\begin{align}
&\frac{\partial k^i}{\partial\phi}\,\frac{\partial k_i}{\partial u}=0\ ,& &\frac{\partial k^i}{\partial\theta}\,\frac{\partial k_i}{\partial u}=-A\,\sin\theta\ ,& \nonumber \\
&v_i\,\frac{\partial k^i}{\partial u}=-A\,\cos\theta\ ,&  &\frac{\partial k^i}{\partial u}\frac{\partial k_i}{\partial u}=-A^2\sin^2\theta\ .& \label{12}
\end{align}
The formula (\ref{9}) determines $\theta, \phi, r, u$ implicitly as functions of the coordinates $X^i=(X, Y, Z, T)$. Hence differentiating (\ref{9}) partially with respect to $X^j$ gives
\begin{equation}\label{13}
\delta^i_j=\left (v^i+r\,\frac{\partial k^i}{\partial u}\right )u_{,j}+k^i\,r_{,j}+r\,\frac{\partial k^i}{\partial\theta}\theta_{,j}+r\,\frac{\partial k^i}{\partial\phi}\phi_{,j}\ .
\end{equation}
with the comma denoting partial differentiation here. From this we have
\begin{equation}\label{14}
dX^i=\left (v^i+r\,\frac{\partial k^i}{\partial u}\right )du+k^i\,dr+r\,\frac{\partial k^i}{\partial\theta}d\theta+r\,\frac{\partial k^i}{\partial\phi}d\phi\ ,
\end{equation}
from which we derive the Minkowskian line element in coordinates $x^{i'}=(\theta, \phi, r, u)$:
\begin{eqnarray}
ds^2&=&-r^2\{(d\theta+A(u)\sin\theta du)^2+\sin^2\theta\,d\phi^2\}+2\,du\,dr \nonumber \\
&&+\left (1-2\,A(u)r\cos\theta\right )\,du^2\ .\label{15}
\end{eqnarray}
Multiplying (\ref{13}) successively by $k_i, v_i, \partial k_i/\partial\theta$ and $\partial k_i/\partial\phi$ results respectively in 
\begin{equation}\label{16}
u_{,j}=k_j\ , r_{,j}=v_j-(1-r\,A\,\cos\theta)k_j\ ,
\end{equation}
and
\begin{equation}\label{17}
\theta_{,j}=-\frac{1}{r}\frac{\partial k_j}{\partial\theta}-A\,\sin\theta\,k_j\ , \phi_{,j}=-\frac{1}{r\sin^2\theta}\frac{\partial k_j}{\partial\phi}\ .
\end{equation}
When these are substituted into (\ref{1}) and the lower index is raised using $\eta^{jk}$ (where $\eta^{jk}\eta_{ki}=\delta ^j_i$) we obtain the useful formula
\begin{equation}\label{18}
\eta^{ij}=-\frac{\partial k^i}{\partial\theta}\frac{\partial k^j}{\partial\theta}-\frac{1}{\sin^2\theta}\frac{\partial k^i}{\partial\phi}\frac{\partial k^j}{\partial\phi}+k^i\,v^j+k^j\,v^i-k^i\,k^j\ ,
\end{equation}
after making use of the identity
\begin{equation}\label{19}
\frac{\partial k^i}{\partial u}-A\,\sin\theta\frac{\partial k^i}{\partial\theta}+A\,\cos\theta\,k^i=0\ .
\end{equation}
The line element of Minkowskian space--time (\ref{15}) can be written in terms of basis 1--forms $\vartheta^{(1)}, \vartheta^{(2)}, \vartheta^{(3)}, \vartheta^{(4)}$ as
\begin{eqnarray}
ds^2&=&\eta_{ij}\,dX^i\,dX^j=-(\vartheta^{(1)})^2-(\vartheta^{(2)})^2+2\,\vartheta^{(3)}\,\vartheta^{(4)} \nonumber \\
&=&g_{(a)(b)}\,\vartheta^{(a)}\,\vartheta^{(b)}\ ,\label{20}
\end{eqnarray}
with
\begin{eqnarray}
\vartheta^{(1)}&=&r\,(d\theta+A\,\sin\theta\,du)=-\vartheta_{(1)}\ ,\nonumber\\
\vartheta^{(2)}&=&r\,\sin\theta\,d\phi=-\vartheta_{(2)}\ ,\nonumber\\
\vartheta^{(3)}&=&dr+\left(\frac{1}{2}-r\,A\,\cos\theta\right )du=\vartheta_{(4)}\ , \nonumber\\
\vartheta^{(4)}&=&du=\vartheta_{(3)}\ .\label{21}
\end{eqnarray}
The 1--forms define a half null tetrad and the tetrad components of the metric tensor are given by $g_{(a)(b)}$ in (\ref{20}). We have used $g_{(a)(b)}$ to lower the tetrad indices in each of (\ref{21}) and we shall raise the tetrad indices using $g^{(a)(b)}$ defined by $g^{(a)(b)}g_{(b)( c )}=\delta^a_c$. Using (\ref{16}) and (\ref{17}) we arrive at
\begin{equation}\label{22}
\vartheta_{(1)}^i=\frac{\partial k^i}{\partial\theta}\ ,\ \vartheta_{(2)}^i=\frac{1}{\sin\theta}\frac{\partial k^i}{\partial\phi}\ ,\ \vartheta_{(3)}^i=k^i\ ,\ \vartheta_{(4)}^i=v^i-\frac{1}{2}\,k^i .
\end{equation}

Having prepared the background space--time in the neighborhood of the world line $r=0$ we now consider the background space--time in more generality. A general axially symmetric form of the line element which incorporates (\ref{15}) as a special case is given by
\begin{eqnarray}
ds^2&=&-r^2p^{-2}\{(e^{\alpha}d\theta+a\,du)^2+e^{-2\,\alpha}\sin^2\theta\,d\phi^2\}\nonumber \\
&&+2\,du\,dr+c\,du^2=g_{i'j'}\,dx^{i'}dx^{j'}\ ,\label{23}
\end{eqnarray}
where the functions $p, \alpha, a$ and $c$ are functions of $\theta, r, u$. This is an axially symmetric special case of a form of the most general line element \cite{Hogan:Trautman:1987} (involving six functions of four coordinates) originally used for studying gravitational radiation from isolated sources. The 3--surfaces $u={\rm constant}$ are null hypersurfaces as in the special case of (\ref{15}). The coordinates $\theta, \phi$ label the null geodesic generators of these hypersurfaces, while $r$ is an affine parameter along the generators. The form of line element (\ref{23}) incorporates the Robinson--Trautman form \cite{Robinson:Trautman:1962} which corresponds to the special case $\alpha=0$. For small values of $r$, and to incorporate (\ref{15}), we shall assume that the functions $p, \alpha, a, c$ can be expanded in powers of $r$ as follows:
\begin{eqnarray}
p&=&1+q_2\,r^2+q_3\,r^3+\dots\ ,\label{24}\\
\alpha&=&\alpha_2\,r^2+\alpha_3\,r^3+\dots\ ,\label{25}\\
a&=&A(u)\,\sin\theta+a_1\,r+a_2\,r^2+\dots\ ,\label{26}\\
c&=&1-2\,r\,A(u)\,\cos\theta+c_2\,r^2+\dots\ ,\label{27}
\end{eqnarray}
with the coefficients of the powers of $r$ functions of $(\theta, u)$. For an axially symmetric Maxwell field we start by choosing a potential 1--form of the form
\begin{equation}\label{28}
{\cal A}=L\,d\theta+K\,du\ ,\end{equation}
with $L$ and $K$ functions of $(\theta, r, u)$. For small positive values of $r$, and in order to arrive at an external Maxwell field which is non--singular on $r=0$, we assume the following expansions of $L$ and $K$ in powers of $r$:
\begin{eqnarray}
L&=&L_2\,r^2+L_3\,r^3+L_4\,r^4+\dots\ ,\nonumber \\
K&=&K_1\,r+K_2\,r^2+K_3\,r^3+\dots\ ,\label{29}
\end{eqnarray}
with the coefficients of the various powers of $r$ functions of $(\theta, u)$. We now write the line element (\ref{23}) in terms of a basis of 1--forms:
\begin{equation}\label{30}
ds^2=-(\vartheta^{(1)})^2-(\vartheta^{(2)})^2+2\,\vartheta^{(3)}\,\vartheta^{(4)}\ ,\end{equation}
with
\begin{eqnarray}
\vartheta^{(1)}&=&r\,p^{-1}(e^{\alpha}d\theta+a\,du)=-\vartheta _{(1)}\ ,\label{31}\\
\vartheta^{(2)}&=&r\,p^{-1}e^{-\alpha}\sin\theta\,d\phi=-\vartheta _{(2)}\ ,\label{32}\\
\vartheta^{(3)}&=&dr+\frac{1}{2}c\,du=\vartheta _{(4)}\ ,\label{33}\\
\vartheta^{(4)}&=&du=\vartheta _{(3)}\ .\label{34}
\end{eqnarray}
The candidate for external Maxwell field is the 2--form
\begin{equation}\label{35}
F=d{\cal A}=\frac{1}{2}F_{(a)(b)}\,\vartheta^{(a)}\wedge\vartheta^{(b)}\ ,\end{equation}
which is the exterior derivative of the 1--form (\ref{28}). With the assumed expansions in powers of $r$ given above we find the following tetrad components of $F$:
\begin{eqnarray}
F_{(1)(2)}&=&0\ , \quad F_{(1)(3)}=-2\,L_2+O( r )\ , \nonumber \\
F_{(1)(4)}&=&L_2+\frac{\partial K_1}{\partial\theta}+O( r )\ , \quad  F_{(2)(3)}=0\ ,\nonumber\\
F_{(2)(4)}&=&0\ , \quad  F_{(3)(4)}=K_1+O( r )\ .\label{36}
\end{eqnarray} 
When these are calculated on $r=0$, and the tetrad vectors (\ref{22}) in coordinates $X^i$ are used, we find that
\begin{eqnarray}
& F_{ij}(u)\,\frac{\partial k^i}{\partial\theta}\frac{\partial k^i}{\partial\phi}=0\ ,\ F_{ij}(u)\,\frac{\partial k^i}{\partial\phi}\,k^j=0\ , \nonumber \\
& F_{ij}(u)\,\frac{\partial k^i}{\partial\phi}\left (v^j-\frac{1}{2}\,k^j\right )=0\ ,\label{37a}
\end{eqnarray}
and
\begin{equation}\label{37b}
K_1=F_{ij}(u)\,k^i\,v^j\ ,\ L_2=-\frac{1}{2}F_{ij}(u)\,\frac{\partial k^i}{\partial\theta}\,k^j\ ,
\end{equation}
with
\begin{equation}\label{37c}
L_2+\frac{\partial K_1}{\partial\theta}=F_{ij}(u)\frac{\partial k^i}{\partial\theta}\left (v^j-\frac{1}{2}\,k^j\right )\ ,
\end{equation}
where $F_{ij}(u)=-F_{ji}(u)$ are the components of the Maxwell tensor in coordinates $X^i$ calculated on the world line $r=0$. We can simplify matters at this point by making the assumption that to sustain rectilinear motion in the $Z$--direction requires only an electric field in the $Z$--direction and so we can satisfy (\ref{37a}) by taking $F_{ij}=0$ except for $F_{34}=-F_{43}=E(u)$ (say). Now with $k^i$ given by (\ref{10}) and $v^i=(0, 0, v^3, v^4)$ we conclude from (\ref{37b}) that
\begin{equation}\label{38}
L_2=-\frac{1}{2}\,E(u)\,\sin\theta\ \ \ {\rm and}\ \ \ K_1=-E(u)\,\cos\theta\ ,
\end{equation}
and thus (\ref{37c}) reduces to 
\begin{equation}\label{39}
L_2+\frac{\partial K_1}{\partial\theta}=\frac{1}{2}\,E(u)\,\sin\theta\ ,
\end{equation}
which is clearly satisfied. Now Maxwell's equations $d{}^*F=0$, with $F$ given by (\ref{35}) and the star indicating the Hodge dual of the 2--form $F$, are satisfied with an $O( r )$--error provided
\begin{equation}\label{40}
\frac{\partial K_1}{\partial\theta}+2\,L_2=0\ .
\end{equation}
This is satisfied by (\ref{38}). Hence we see that the tetrad components of the Maxwell tensor $F_{(a)(b)}=-F_{(b)(a)}$ on $r=0$ vanish except for $F_{(1)(3)}=E(u)\,\sin\theta$, $F_{(1)(4)}=\frac{1}{2}E(u)\,\sin\theta$ and $F_{(3)(4)}=-E(u)\,\cos\theta$. We note for future reference that the components $E_{ij}(u)$ of the electromagnetic energy--momentum tensor calculated on $r=0$ in the coordinates $X^i$ are given by
\begin{eqnarray}
E_{ij}&=&F_{ik}\,F_j{}^k-\frac{1}{4}\eta_{ij}F_{kl}\,F^{kl} \nonumber\\
&=&\frac{1}{2}\,{\rm diag}\left (-E^2, -E^2, +E^2, -E^2\right )\ .\label{41}
\end{eqnarray}
where the indices on $F_{ij}$ are raised using $\eta^{ij}$ with $\eta^{ik}\eta_{kj}=\delta^i_j$.

With (\ref{38}) we have evaluated the leading terms in the expansions (\ref{29}). In the sequel we will require the next term in each of these expansions. In other words we will require the functions $L_3$ and $K_2$. These are given by the vanishing of the next to leading terms in the expansion of Maxwell's equations $d{}^*F=0$ in positive powers of $r$. The results are the differential equations
\begin{equation}\label{q1}
\frac{\partial}{\partial\theta}(L_3\,\sin\theta)-2\,K_2\sin\theta=A\,E\,\sin^3\theta\ ,
\end{equation}
and
\begin{equation}\label{q2}
\sin\theta\,\frac{\partial K_2}{\partial\theta}+3\,L_3\,\sin\theta=-\dot E\,\sin^2\theta-A\,E\,\sin^2\theta\,\cos\theta\ ,
\end{equation}
with $\dot E=dE/du$. From these we easily see that $K_2$ satisfies the inhomogeneous $l=2$ Legendre equation, with the right hand side a linear combination of an $l=0$ and an $l=1$ Legendre polynomial:
\begin{equation}\label{q3}
\frac{1}{\sin\theta}\frac{\partial}{\partial\theta}\left (\sin\theta\,\frac{\partial K_2}{\partial\theta}\right )+6\,K_2=-2\,A\,E-2\,\dot E\,\cos\theta\ .
\end{equation}
The general solution which is non--singular for $0\leq\theta\leq\pi$ is
\begin{equation}\label{q4}
K_2=-\frac{1}{3}A\,E-\frac{1}{2}\dot E\,\cos\theta+\omega(u)\,(3\,\cos^2\theta-1)\ ,
\end{equation}
where $\omega(u)$ is an arbitrary function of integration. The corresponding expression for $L_3$ follows from (\ref{q2}) and is
\begin{equation}\label{q5}
L_3=-\frac{1}{2}\,\dot E\,\sin\theta+\left (2\,\omega-\frac{1}{3}A\,E\right )\sin\theta\,\cos\theta\ .
\end{equation}
In the sequel we will assume that the external electromagnetic field does not involve a second independent arbitrary function of $u$ in addition to $E(u)$ and so we will take $\omega(u)=0$. The non--vanishing tetrad components of the external Maxwell field now read
\begin{eqnarray}
F_{(1)(3)}&=&E\,\sin\theta+\left (\frac{3}{2}\,\dot E+A\,E\,\cos\theta\right )\,r \sin\theta+O(r^2)\ , \nonumber \\ \label{p1}\\
F_{(1)(4)}&=&\frac{1}{2}E\,\sin\theta+\Big(\frac{1}{4}\,\dot E \nonumber\\
&&+\frac{1}{2}\,A\,E\,\cos\theta\Big)\,r \sin\theta+O(r^2)\ , \label{p2}\\
F_{(3)(4)}&=&-E\,\cos\theta-\Big\{\dot E\,\cos\theta \nonumber \\
&&+\frac{1}{3}A\,E\,(3\,\cos^2\theta-1)\Big\}\,r+O(r^2)\ .\label{p3}
\end{eqnarray}
The components $F_{ij}(u)=-F_{ji}(u)$ of the external Maxwell field calculated on $r=0$ in the coordinates $X^i$ can be recovered from (\ref{p1})--(\ref{p3}) using the formula
\begin{equation}\label{p4}
F_{ij}(u)=F_{(a)(b)}\,\vartheta^{(a)}_i\,\vartheta^{(b)}_j\ ,
\end{equation} 
with $\vartheta^{(a)}_i=g^{(a)(b)}\eta_{ij}\vartheta^j_{(b)}$,  $g^{(a)(b)}$ defined by $g^{(a)(c)}\,g_{(c)(b)}=\delta^{a}_{b}$, $g_{(a)(b)}$ given by (\ref{20}) and $\vartheta^i_{(a)}$ by (\ref{22}). This results in $F_{ij}(u)=0$ except for $F_{34}(u)=-F_{43}(u)=E(u)$ as before.

We now turn our attention to Einstein's field equations
\begin{equation}\label{42}
R_{(a)(b)}=2\,E_{(a)(b)}\ ,
\end{equation}
where $R_{(a)(b)}$ are the components of the Ricci tensor on the tetrad defined by (\ref{31})--(\ref{34}) and $E_{(a)(b)}$ are the components of the electromagnetic energy--momentum tensor on the tetrad. When the expansions (\ref{24})--(\ref{27}) and (\ref{29}) in powers of $r$ are introduced this results in $R_{(a)(b)}-2\,E_{(a)(b)}$ having expansions in powers of $r$ starting with terms independent of $r$. We will only require the functions $q_2, \alpha_2, a_1, c_2$ appearing in the expansions (\ref{24})--(\ref{27}) and these can be obtained by requiring the vanishing of the terms independent of $r$ in $R_{(a)(b)}-2\,E_{(a)(b)}$. This results in the following seven equations to be satisfied by  $q_2, \alpha_2, a_1, c_2$:
\begin{eqnarray}
&&q_2=\frac{1}{6}\,E^2\sin^2\theta\ ,\label{43}\\
&&\frac{\partial}{\partial\theta}(q_2+\alpha_2)+2\,\alpha_2\,\cot\theta-a_1=E^2\sin\theta\cos\theta\ ,\label{44}\\
&&\frac{\partial}{\partial\theta}(\alpha_2+q_2-c_2)+2\,\alpha_2\,\cot\theta+2\,a_1=E^2\sin\theta\cos\theta\ , \nonumber \\\label{45}\\
&&\frac{\partial a_1}{\partial\theta}+a_1\,\cot\theta+2\,c_2-4\,q_2=\frac{2}{3}E^2\cos^2\theta\ ,\label{46}\\
&&\frac{3}{2}\left (\frac{\partial a_1}{\partial\theta}+a_1\,\cot\theta\right )-\frac{1}{2\,\sin\theta}\frac{\partial}{\partial\theta}\left (\sin\theta\frac{\partial c_2}{\partial\theta}\right )-3\,q_2 \nonumber \\
&&\quad =-\frac{1}{2}E^2\sin^2\theta\ ,\label{47}\\
&&4\,\frac{\partial a_1}{\partial\theta}+a_1\,\cot\theta-\chi+3\,c_2+6\,\alpha_2-12\,q_2\nonumber \\
&&\quad=E^2-2\,E^2\cos^2\theta\ ,\label{48}\\
&&\frac{\partial a_1}{\partial\theta}+4\,a_1\,\cot\theta-\chi+3\,c_2-6\,\alpha_2-12\,q_2=-E^2\ ,\label{49}
\end{eqnarray}
with
\begin{equation}\label{50}
\chi=\frac{1}{\sin\theta}\frac{\partial}{\partial\theta}\left\{\sin\theta\frac{\partial}{\partial\theta}(q_2+\alpha_2)+2\,\alpha_2\cos\theta\right\}\ .
\end{equation}
Combining (\ref{46}) and (\ref{47}), and using $q_2$ given by (\ref{43}), we arrive at 
\begin{equation}\label{51}
\frac{1}{\sin\theta}\frac{\partial}{\partial\theta}\left (\sin\theta\frac{\partial c_2}{\partial\theta}\right )+6\,c_2=2\,E^2\ ,
\end{equation}
which is the inhomogeneous $l=2$ Legendre equation with an $l=0$ Legendre polynomial on the right hand side. The general solution of this equation which is non--singular for $0\leq\theta\leq\pi$ is  
\begin{equation}\label{52}
c_2(\theta, u)=\frac{1}{3}E^2+C(u)(3\,\cos^2\theta-1)\ .
\end{equation}
Here $C(u)$ is an arbitrary function of integration. We demonstrate below that $C(u)$ is simply related to the Weyl tensor of the space--time on $r=0$ in coordinates $X^i$. Knowing $q_2$ and $c_2$ we see that (\ref{46}) provides us with an equation for $a_1$, namely,
\begin{equation}\label{53}
\frac{\partial}{\partial\theta}(a_1\,\sin\theta)=2\,C(u)\,(3\,\cos^2\theta-1)\sin\theta\ ,
\end{equation}
and the solution which is non--singular for $0\leq\theta\leq\pi$ is
\begin{equation}\label{54}
a_1(\theta, u)=-2\,C\,\sin\theta\cos\theta\ .
\end{equation}
Now (\ref{44}) becomes
\begin{equation}\label{55}
\frac{\partial}{\partial\theta}(\alpha_2\,\sin^2\theta)=\left (-2\,C+\frac{2}{3}E^2\right )\sin^3\theta\cos\theta\ ,
\end{equation}
and the solution of this equation which is non--singular for $0\leq\theta\leq\pi$ is
\begin{equation}\label{56}
\alpha_2(\theta, u)=\left (-\frac{1}{2}C+\frac{1}{6}E^2\right )\sin^2\theta\ .
\end{equation}
It is now straightforward to check that $q_2, c_2, a_1, \alpha_2$ given by (\ref{43}), (\ref{52}), (\ref{54}) and (\ref{56}) respectively satisfy (\ref{44})--(\ref{49}).
 
The approximate solution of the equations (\ref{42}) given above involves two arbitrary functions of $u$, namely $E(u)$ and $C(u)$. We have already seen that $E(u)=F_{34}(u)$ where $F_{34}(u)$ is the non--vanishing component of the external Maxwell field calculated on $r=0$ in coordinates $X^i$. We now seek to demonstrate how $C(u)$ is related to the components $C_{ijkl}(u)$ of the Weyl tensor of the space--time (the external gravitational field) calculated on $r=0$ in coordinates $X^i$. We start by listing in the Appendix \ref{app_formulas} the tetrad components of the Riemann tensor calculated on $r=0$ using the functions $q_2, c_2, a_1, \alpha_2$ derived above. The components $R_{ijkl}(u)$ of the Riemann tensor calculated on $r=0$ in the coordinates $X^i$ are given in terms of these tetrad components by
\begin{equation}\label{69}
R_{ijkl}(u)=R_{(a)(b)( c )(d)}\,\vartheta^{(a)}_i\,\vartheta^{(b)}_j\,\vartheta^{( c )}_k\,\vartheta^{(d)}_l\ ,
\end{equation}
with $\vartheta^{(a)}_i$ given by (\ref{22}) (with $\vartheta^{(a)}_i=g^{(a)(b)}\eta_{ij}\vartheta^j_{(b)}$) since we are calculating on $r=0$ in coordinates $X^i$. We find that the non--vanishing $R_{ijkl}(u)$ are  
\begin{eqnarray}
R_{1212}(u)&=&-\frac{5}{3}E^2+2\,C\ ,\label{70}\\
R_{1313}(u)&=&R_{2323}(u)=\frac{1}{3}E^2-C\ ,\label{71}\\
R_{1414}(u)&=&R_{2424}(u)=-\frac{1}{3}E^2+C\ ,\label{72}\\
R_{3434}(u)&=&-\frac{1}{3}E^2-2\,C\ .\label{73}
\end{eqnarray}
A useful check on these components is to verify that 
\begin{equation}\label{74}
\eta^{kl}\,R_{kijl}(u)=R_{ij}(u)=2\,E_{ij}(u)\ ,
\end{equation}
with $E_{ij}(u)$ given by (\ref{41}) confirming that Einstein's field equations are satisfied on $r=0$. Taking into consideration (\ref{74}) the components of the Weyl tensor on $r=0$ in coordinates $X^i$ are 
\begin{eqnarray}
C_{ijkl}(u)&=&R_{ijkl}(u)+\eta_{ik}\,E_{jl}(u)-\eta_{il}\,E_{jk}(u) \nonumber \\
&+&\eta_{jl}\,E_{ik}(u)-\eta_{jk}\,E_{il}(u)\ .\label{75}
\end{eqnarray}
Using (\ref{41}) and (\ref{70})--(\ref{73}) we find that the non--vanishing components of the Weyl tensor on $r=0$ in coordinates $X^i$ are 
\begin{eqnarray}
\frac{1}{2}C_{1212}=-\frac{1}{2}C_{3434}=-C_{1313}=-C_{2323} \nonumber \\
=C_{1414}=C_{2424}=-\frac{1}{3}E^2+C\ .\label{76}
\end{eqnarray}
This determines the relationship between the arbitrary function $C(u)$ and the external gravitational field or Weyl tensor. In the sequel we shall assume, for simplicity, that the particle with world line $r=0$ experiences only an electric field as external field. Thus the external field involves only one arbitrary function $E(u)$ and we shall take $C(u)=0$ from now on leading to the simplifications
\begin{equation}\label{76a}
\alpha_2=\frac{1}{6}E^2\sin^2\theta\ ,\ \ a_1=0\ \ {\rm and}\ \ c_2=\frac{1}{3}E^2\sin^2\theta\ ,
\end{equation}
with $q_2$ given by (\ref{43}).

\section{A Charged Mass Particle}\label{sec_charged_mass_particle}

We introduce a particle of small mass $m=O_1$ and small charge $e=O_1$ (both constant) as a perturbation of the potential 1--form and space--time above by modifying (\ref{28}) and (\ref{30}) using the expansions:
\begin{equation}\label{77}
K=\frac{(e+\hat K_{-1})}{r}+\hat K_0+\hat K_1\,r+\hat K_2\,r^2+\dots\ ,
\end{equation}
with $\hat K_{-1}=O_2\ ,\hat K_0=O_1\ ,\ \hat K_1=-E\,\cos\theta+O_1$ and $\hat K_2=-\frac{1}{3}A\,E-\frac{1}{2}\dot E\,\cos\theta+O_1$;
\begin{equation}\label{78}
L=\hat L_2\,r^2+\hat L_3\,r^3+\dots\ ,
\end{equation}
with $\hat L_2=-\frac{1}{2}\,E\,\sin\theta+l_2+O_2$, $l_2=O_1$ and $\hat L_3=-\frac{1}{2}\,\dot E\,\sin\theta-\frac{1}{3}A\,E\,\sin\theta\,\cos\theta+O_1$;
\begin{equation}\label{79}
p=\hat P(1+\hat q_2\,r^2+\dots\ )\ ,
\end{equation}
with $\hat P=1+Q_1+Q_2+O_3$ and $Q_1=O_1, Q_2=O_2$ and $\hat q_2=\frac{1}{6}\,E^2\sin^2\theta+O_1$;
\begin{equation}\label{80}
\alpha=\hat\alpha_2\,r^2+\dots\ ,
\end{equation}
with $\hat\alpha_2=\frac{1}{6}E^2\sin^2\theta+O_1$;
\begin{equation}\label{81}
a=\frac{\hat a_{-1}}{r}+\hat a_0+\hat a_1\,r+\dots\ ,
\end{equation}
with $\hat a_{-1}=O_2\ , \hat a_0=A\,\sin\theta+O_1\ , \hat a_1=O_1$;
\begin{equation}\label{82}
c=\frac{e^2}{r^2}-\frac{2\,(m+\hat c_{-1})}{r}+\hat c_0+\hat c_1\,r+\hat c_2\,r^2+\dots\ ,
\end{equation}
with $e^2=O_2\ , m=O_1\ , \hat c_{-1}=O_2\ , \hat c_0=1+O_1\ , \hat c_1=-2\,A\,\cos\theta+O_1$ and $\hat c_2=\frac{1}{3}E^2+O_1$. These expansions ensure that for small values of $r$ the perturbed field (gravitational and electromagnetic) resembles the Reissner--Nordstr\"om field. The exact Maxwell 2--form (\ref{35}) reads
\begin{equation}\label{83}
F=\frac{\partial L}{\partial r}\,dr\wedge d\theta+\left (\frac{\partial K}{\partial\theta}-\frac{\partial L}{\partial u}\right )d\theta\wedge du+\frac{\partial K}{\partial r}dr\wedge du\ ,
\end{equation}
and its Hodge dual is
\begin{equation}\label{84}
{}^*F=F_1\,dr\wedge d\phi+F_2\,d\phi\wedge du+F_3\,d\theta\wedge d\phi\ ,
\end{equation}
with
\begin{eqnarray}
F_1&=&-e^{-2\alpha}\sin\theta\,\frac{\partial L}{\partial r}\ ,\label{85}\\
F_2&=&c\,e^{-2\alpha}\sin\theta\,\frac{\partial L}{\partial r}+e^{-2\alpha}\sin\theta\left (\frac{\partial K}{\partial\theta}-\frac{\partial L}{\partial u}\right )\nonumber\\
&&+a\,r^2p^{-2}e^{-\alpha}\sin\theta\left (\frac{\partial K}{\partial r}-a\,e^{-\alpha}\frac{\partial L}{\partial r}\right )\ ,\label{86}\\
F_3&=&r^2p^{-2}\sin\theta\left (a\,e^{-\alpha}\frac{\partial L}{\partial r}-\frac{\partial K}{\partial r}\right )\ .\label{87}
\end{eqnarray}
In terms of $F_1, F_2, F_3$ the exact Maxwell equations read
\begin{equation}\label{88}
\frac{\partial F_3}{\partial r}-\frac{\partial F_1}{\partial\theta}=0\ ,\ 
\frac{\partial F_2}{\partial\theta}+\frac{\partial F_3}{\partial u}=0\ , \ \frac{\partial F_1}{\partial u}+\frac{\partial F_2}{\partial r}=0\ .
\end{equation}
We note that in coordinates $x^{i'}=(\theta, \phi, r, u)$ with $g={\rm det}(g_{i'j'})$, if $F^{i'j'}=-F^{j'i'}$ are the components of the Maxwell tensor, then $F^{2'j'}\equiv0$ and (\ref{88}) are equivalent to 
\begin{equation}\label{88'}
\sqrt{-g}\,F^{4'j'}{}_{;j'}=0\ ,\  \sqrt{-g}\,F^{3'j'}{}_{;j'}=0\ ,\ \sqrt{-g}\,F^{1'j'}{}_{;j'}=0\ ,
\end{equation}
respectively, with the semicolon denoting covariant differentiation with respect to the Riemannian connection calculated with a metric tensor $g_{i'j'}$ given via (\ref{23}). In order to solve (\ref{88}) approximately we first substitute the expansions (\ref{77})--(\ref{82}) into the functions $F_1, F_2, F_3$ to arrive at
\begin{eqnarray}
F_1&=&\left (E(u)\,\sin^2\theta-2\,l_2\,\sin\theta+O_2\right )r+\biggl\{\frac{3}{2}\dot E\,\sin^2\theta\nonumber\\
&&+A\,E\,\sin^2\theta\,\cos\theta+O_1\biggr\}\,r^2+O(r^3)\ ,\label{89}\\
F_2&=&\frac{\sin\theta}{r}\left\{\frac{\partial\hat K_{-1}}{\partial\theta}-e^2E\,\sin\theta-e\,\hat a_{-1}+O_3\right\}\nonumber\\
&&+\sin\theta\biggl\{\frac{\partial\hat K_0}{\partial\theta}+2\,m\,E(u)\,\sin\theta-e\,\hat a_0\nonumber\\
&&-\frac{3}{2}e^2\dot E\,\sin\theta+2\,e\,A\,Q_1\,\sin\theta-e^2A\,E\,\sin\theta\,\cos\theta\nonumber\\
&&-4\,m\,l_2+2\,E\,\hat c_{-1}\,\sin\theta-A\,\hat K_{-1}\,\sin\theta+O_3\biggr\}\nonumber\\
&&+r\,\sin\theta\biggl\{3\,m\,\dot E\,\sin\theta+2\,m\,A\,E\,\sin\theta\,\cos\theta \nonumber \\
&&-\hat c_0\,E\,\sin\theta+2\,l_2 +\frac{\partial\hat K_1}{\partial\theta} \nonumber \\
&&+\hat a_{-1}\,\hat K_1+O_2\biggr\}+O(r^2)\ ,\label{90}
\end{eqnarray}
\begin{eqnarray}
F_3&=&e\,\sin\theta-2\,e\,Q_1\,\sin\theta+\hat K_{-1}\sin\theta+O_3 \nonumber \\
&&-\Biggl\{\hat K_1+\hat a_{-1}E\,\sin\theta+2\,Q_1\,E\,\cos\theta\nonumber\\
&&+\frac{1}{3}e\,E^2\sin^2\theta+O_2\Biggr\}r^2\sin\theta+\biggl\{\dot E\,\cos\theta\nonumber\\
&&+\frac{1}{3}\,A\,E\,(3\,\cos^2\theta-1)+O_1\biggr\}\,r^3\sin\theta+O(r^4)\ . \nonumber \\\label{91}\end{eqnarray}
Now (\ref{88}) are satisfied approximately in the sense that
\begin{eqnarray}
\frac{\partial F_3}{\partial r}-\frac{\partial F_1}{\partial\theta}&=&O_1\times r+O(r^2)\ ,\label{92}\\
\frac{\partial F_2}{\partial\theta}+\frac{\partial F_3}{\partial u}&=&O_3\times\frac{1}{r}+O_2+O_1\times r+O(r^2)\ ,\label{93}\\
\frac{\partial F_1}{\partial u}+\frac{\partial F_2}{\partial r}&=&O_3\times\frac{1}{r^2}+O_1+O(r)\ ,\label{94}\end{eqnarray}
provided 
\begin{eqnarray}
\hat K_1&=&-E\,\cos\theta+\frac{1}{\sin\theta}\frac{\partial}{\partial\theta}(l_2\sin\theta)-\hat a_{-1}E\,\sin\theta\nonumber\\
&&-\frac{1}{3}e\,E^2\sin^2\theta-2\,E\,Q_1\cos\theta+O_2\ ,\label{95''} \nonumber\\
\frac{\partial\hat K_{-1}}{\partial\theta}&=&e^2E\,\sin\theta+e\,\hat a_{-1}\ ,\label{95'}
\end{eqnarray}
and $\hat K_0$ satisfies
\begin{equation}\label{95}
\sin\theta\left\{\frac{\partial\hat K_0}{\partial\theta}+\left (2\,m\,E(u)-e\,A(u)\right )\sin\theta+O_2\right \}=U(u)\ ,
\end{equation}
for some function $U(u)$ of integration. For a Maxwell field which is non--singular for $0\leq\theta\leq\pi$ we must have $U(u)=0$ and thence
\begin{equation}\label{96}
\hat K_0=\left (2\,m\,E(u)-e\,A(u)\right )\cos\theta+O_2\ .
\end{equation}
A function of $u$ of integration added to this is easily seen to be a pure gauge term in the potential 1--form and so it has been neglected. The second term in (\ref{96}) is the Li\'enard--Wiechert contribution to the potential 1--form (\ref{28}). We will return to $F_2$ in section \ref{sec_res_matter_dist} and in particular determine the $O_2$--part of $\hat K_0$ by requiring (\ref{93}) to read
\begin{equation}\label{966}
\frac{\partial F_2}{\partial\theta}+\frac{\partial F_3}{\partial u}=O_3\times\frac{1}{r}+O_3+O_1\times r+O(r^2)\ .\end{equation}
To solve (\ref{95'}) for $\hat K_{-1}$ we must first obtain $\hat a_{-1}$ by solving approximately one of the Einstein field equations $R_{i'j'}-2\,E_{i'j'}=0$ in coordinates $x^{i'}=(\theta, \phi, r, u)$ with $i'=1', 2', 3', 4'$. Specifically, requiring the coefficient of $r^{-1}$ in $R_{1'3'}-2\,E_{1'3'}$ to be small of second order yields
\begin{equation}\label{97}
\hat a_{-1}=2\,e\,E\,\sin\theta+O_2\ ,
\end{equation}
and combining this with (\ref{95'}) results in
\begin{equation}\label{98}
\hat K_{-1}=-3\,e^2E\,\cos\theta+k(u)+O_3\ \ \ {\rm with}\ \ k(u)=O_2\ ,
\end{equation}
where $k(u)$ is an arbitrary function of integration.

With $\hat a_{-1}$ given by (\ref{97}) and $\hat P$ given by (\ref{79}) we find that requiring the coefficient of $r^{0}$ in $R_{2'2'}-2\,E_{2'2'}$ to be small of second order provides us with
\begin{equation}\label{99}
\hat c_0=1+\underset{0}{\Delta}Q_1+2\,Q_1-8\,e\,E\,\cos\theta+O_2\ ,
\end{equation}
where 
\begin{equation}\label{99'}
\underset{0}{\Delta}Q_1=\frac{1}{\sin\theta}\frac{\partial}{\partial\theta}\left (\sin\theta\frac{\partial Q_1}{\partial\theta}\right )\ .
\end{equation}
If we now require the coefficient of $r^{-2}$ in $R_{4'4'}-2\,E_{4'4'}$ to be small of second order we find that 
\begin{equation}\label{100}
\frac{1}{2}\underset{0}{\Delta}\hat c_0=(2\,e\,E-6\,m\,A)\,\cos\theta+O_2\ .
\end{equation}
On substituting from (\ref{99}) into this
\begin{equation}\label{101}
-\frac{1}{2}\underset{0}{\Delta}(\underset{0}{\Delta}Q_1+2\,Q_1)=6(m\,A+e\,E)\cos\theta+O_2\ .
\end{equation}
This has the general solution, which is non--singular for $0\leq\theta\leq\pi$,
\begin{equation}\label{102}
\underset{0}{\Delta}Q_1+2\,Q_1=6(m\,A+e\,E)\cos\theta+\varphi(u)+O_2\ ,
\end{equation}
where $\varphi(u)=O_1$ is an arbitrary function of integration. This equation has the particular integral $-(m\,A+e\,E)\cos\theta\,\log(\sin^2\theta)$ which is singular when $\theta=0$ or $\pi$ so to have a solution $Q_1$ of (\ref{102}) which is non--singular for $0\leq\theta\leq\pi$ we must have
\begin{equation}\label{103}
m\,A=-e\,E+O_2\,
\end{equation}
Now the general non--singular solution of (\ref{102}) is
\begin{equation}\label{104}
Q_1=\frac{1}{2}\varphi(u)+\psi(u)\,\cos\theta\ ,
\end{equation}
where $\psi(u)=O_1$ is another arbitrary function of integration. This is a linear combination of an $l=0$ and an $l=1$ Legendre polynomial. The 2--surfaces $u={\rm constant}$, $r={\rm constant}$, for small values of $r$, have line elements (specializing (\ref{23}))
\begin{equation}\label{105}
dl^2=r^2(1-2\,Q_1+O_2)(d\theta^2+\sin^2\theta\,d\phi^2)+O(r^4)\ .
\end{equation}
These 2--surfaces play the role of the wave fronts of the radiation produced by the motion of the particle having mass $m$ and charge $e$. Near the particle (for small $r$) these 2--surfaces are smooth perturbations of 2--spheres. However it is well--known (see, for example, \cite{Ferraro:1962}) that perturbations in which $Q_1$ is an $l=0$ or $l=1$ Legendre polynomial are trivial in the sense that the ``perturbed" 2--sphere remains a 2--sphere in these cases. We will discard such perturbations (by putting $\varphi=0$ and $\psi=0$ in (\ref{104})) and so take $Q_1=0$ with (\ref{103}) holding. We note that $E(u)=F_{34}(u)=-F_{43}(u)$ where $F_{34}$ is the non--vanishing component of the external Maxwell tensor in coordinates $X^i$, calculated on the world line $r=0$ in the background space--time described in section \ref{sec_external_fields}. As pointed out prior to (\ref{9}) the non--vanishing components of the 4--acceleration $a^i$ of the particle satisfy $a^3=A\,v^4$ and $a^4=A\,v^3$. Thus since
\begin{equation}\label{106}
E=F_{34}=-F_{43}=-F^3{}_4=-F^4{}_3\ ,
\end{equation}
we have, on account of (\ref{103}),
\begin{equation}\label{107}
m\,a^3=m\,A\,v^4=-e\,E\,v^4+O_2=e\,F^3{}_4\,v^4+O_2\ ,
\end{equation}
and
\begin{equation}\label{107'}
m\,a^4=m\,A\,v^3=-e\,E\,v^3+O_2=e\,F^4{}_3\,v^3+O_2\ ,
\end{equation}
confirming the appearance of the Lorentz 4--force on the right hand side of these equations. 

With $Q_1=0$ and $\hat a_{-1}$ given by (\ref{97}) we see from (\ref{95''}) that now
\begin{equation}\label{108}
\hat K_1=-E\,\cos\theta+\frac{1}{\sin\theta}\frac{\partial}{\partial\theta}(l_2\sin\theta)-\frac{7}{3}e\,E^2\sin^2\theta+O_2\ .
\end{equation}
The coefficient of $r^{-1}$ in $R_{1'3'}-2\,E_{1'3'}$ was required to be small of second order to obtain (\ref{97}). If we now require it to be small of third order we get the more accurate result
\begin{eqnarray}
\hat a_{-1}&=&-4e(L_2+l_2)-4\,\hat K_{-1}\,L_2+O_3\ \nonumber\\
&=&2\,e\,E\,\sin\theta-4\,e\,l_2+2\,k\,E\,\sin\theta \nonumber \\
&&-6\,e^2E^2\sin\theta\,\cos\theta+O_3\ ,\label{109}
\end{eqnarray}
using $L_2$ given by (\ref{78}) and $\hat K_{-1}$ given by (\ref{98}). With our assumptions the quantities $R_{1'1'}-2\,E_{1'1'}$ and $R_{2'2'}-2\,E_{2'2'}$ both have the form $O_3\times r^{-2}+O(r^0)$. We can reduce this form to $O_3\times r^{-2}+O_3+O( r )$ in each of these cases by having respectively
\begin{eqnarray}
\hat c_0&=&1-8\,e\,E\,\cos\theta+\underset{0}{\Delta}Q_2+2\,Q_2+10\,e\,\frac{\partial l_2}{\partial\theta}+6\,e\,l_2\cot\theta\nonumber\\
&&-8\,k\,E\,\cos\theta+24\,e^2E^2-33\,e^2E^2\sin^2\theta+O_3\ ,\label{110}
\end{eqnarray}
and
\begin{eqnarray}
\hat c_0&=&1-8\,e\,E\,\cos\theta+\underset{0}{\Delta}Q_2+2\,Q_2+6\,e\,\frac{\partial l_2}{\partial\theta}+10\,e\,l_2\cot\theta\nonumber\\
&&-8\,k\,E\,\cos\theta+24\,e^2E^2-\frac{103}{3}e^2E^2\sin^2\theta+O_3,\label{111}
\end{eqnarray}
with $\underset{0}{\Delta}$ the operator defined by (\ref{99'}). Subtracting these we find that
\begin{equation}\label{112}
l_2=\frac{1}{3}e\,E^2\sin\theta\,\cos\theta+U(u)\,\sin\theta+O_3\ ,
\end{equation}
where $U(u)=O_1$ is a function of integration. Now (\ref{112}) substituted into (\ref{110}) or (\ref{111}) yields
\begin{eqnarray}
\hat c_0&&=1+\underset{0}{\Delta}Q_2+2\,Q_2+\frac{14}{9}e^2E^2+\biggl\{-8\,e\,E+16\,e\,U\nonumber\\
&&-8\,k\,E\biggr\}\cos\theta+\frac{125}{9}e^2E^2(3\,\cos^2\theta-1)+O_3\ .\label{113}\end{eqnarray}
The final three terms here are a linear combination of $l=0$, $l=1$ and $l=2$ Legendre polynomials. With $l_2$ given by (\ref{112}) we find from (\ref{108}) and (\ref{109}) that 
\begin{equation}\label{114}
\hat K_1=-E\,\cos\theta+2\,U\,\cos\theta-\frac{8}{3}e\,E^2+\frac{10}{3}e\,E^2\cos^2\theta+O_2\ ,
\end{equation}
and
\begin{eqnarray}
\hat a_{-1}&=&2\,e\,E\sin\theta-4\,e\,U\,\sin\theta+2\,k\,E\,\sin\theta \nonumber\\
&&-\frac{22}{3}e^2E^2\sin\theta\cos\theta+O_3\ .\label{115}
\end{eqnarray}
Furthermore requiring $R_{1'1'}-2\,E_{1'1'}$ and $R_{2'2'}-2\,E_{2'2'}$ to both have the form $O_3\times r^{-2}+O_3+O_2\times r+O(r^2)$ we arrive at the two equations
\begin{eqnarray}
\hat c_1&=&2\,e\,A\,E\,\sin^2\theta-\frac{2}{3}e\,A\,E-e\,\dot E\,\cos\theta-\frac{5}{3}m\,E^2\sin^2\theta\nonumber\\
&&-\frac{3}{2}\frac{\partial\hat a_0}{\partial\theta}-\frac{1}{2}\hat a_0\,\cot\theta+O_2\ ,\label{116}
\end{eqnarray}
and
\begin{eqnarray}
\hat c_1&=&-\frac{2}{3}e\,A\,E-e\,\dot E\,\cos\theta-m\,E^2\sin^2\theta-\frac{1}{2}\frac{\partial\hat a_0}{\partial\theta}\nonumber\\
&&-\frac{3}{2}\hat a_0\,\cot\theta+O_2\ .\label{117}
\end{eqnarray}
Subtracting these, remembering that $\hat a_0=A\,\sin\theta+O_1$, we find that
\begin{eqnarray}
\hat a_0&=&A\,\sin\theta+\left\{\frac{2}{3}m\,E^2-2\,e\,A\,E\right\}\sin\theta\cos\theta\nonumber\\
&&+V(u)\,\sin\theta+O_2\ ,\label{118}
\end{eqnarray}
with the function of integration $V(u)=O_1$. With this information we have from (\ref{116}) or (\ref{117}) that
\begin{eqnarray}
\hat c_1&&=-\frac{8}{9}m\,E^2-\left (2\,A+e\,\dot E+2\,V\right )\cos\theta\nonumber\\
&&+\left (\frac{5}{3}e\,A\,E-\frac{2}{9}m\,E^2\right )(3\,\cos^2\theta-1)+O_2\ ,\label{119}
\end{eqnarray}
and we note the appearance of the Legendre polynomials of degree 0, 1 and 2. Next we consider $R_{4'4'}-2\,E_{4'4'}$. One finds directly that this component has the form $O_5\times r^{-6}+O_4\times r^{-5}+O(r^{-4})$. However the coefficient of $r^{-4}$ is $-2\,e\,k+O_4$ with $k(u)$ the $O_2$--function of integration which first appeared in (\ref{98}). We shall therefore take $k(u)=0$ so the $R_{4'4'}-2\,E_{4'4'}$ has the more accurate form $O_5\times r^{-6}+O_4\times r^{-5}+O_4\times r^{-4}+O(r^{-3})$. The coefficient of $r^{-3}$ here is $O_3$ if
\begin{equation}\label{120}
\underset{0}{\Delta}\hat c_{-1}=(8\,m\,e\,E-4\,e^2A)\cos\theta+O_3\ .
\end{equation}
The general solution of this equation which is non--singular for $0\leq\theta\leq\pi$ is
\begin{equation}\label{121}
\hat c_{-1}=(2\,e^2A-4\,m\,e\,E)\cos\theta+g(u)+O_3\ ,
\end{equation}
where $g(u)=O_2$ is an arbitrary function of integration. Now $R_{4'4'}-2\,E_{4'4'}$ has the form $O_5\times r^{-6}+O_4\times r^{-5}+O_4\times r^{-4}+O_3\times r^{-3}+O(r^{-2})$. The coefficient of $r^{-2}$ here is $O_3$ provided 
\begin{eqnarray}
&&-\frac{1}{2}\underset{0}{\Delta}\hat c_0=-\hat a_{-1}\frac{\partial^2\hat a_{-1}}{\partial\theta^2}-2\,\frac{\partial\hat c_{-1}}{\partial u}+m\,\frac{\partial\hat a_0}{\partial\theta}+\hat c_{-1}\frac{\partial\hat a_0}{\partial\theta}\nonumber\\
&&-4\,\frac{\partial\hat K_{-1}}{\partial\theta}\frac{\partial\hat K_1}{\partial\theta}+\frac{1}{2}\hat a_{-1}^2\hat c_0-16\,m^2\,L_2^2-2\,m\,\hat c_1-2\,\hat c_1\hat c_{-1}\nonumber\\
&&+8\,m^2\,q_2-e^2\hat K_1^2-\frac{1}{2}\,\hat a_{-1}\frac{\partial\hat c_0}{\partial\theta}+4\,\hat a_0\,\frac{\partial\hat c_{-1}}{\partial\theta}+3\,e^2\hat c_2\nonumber\\
&&-\left (\frac{\partial\hat a_{-1}}{\partial\theta}\right )^2-2\,\left (\frac{\partial\hat K_0}{\partial\theta}\right )^2+8\,m\,L_2\,\frac{\partial\hat K_0}{\partial\theta}\nonumber\\
&&-4\,\hat c_0\,L_2\,\frac{\partial\hat K_{-1}}{\partial\theta}-4\,m\,\hat a_{-1}\,\hat a_0+4\,e\,\hat a_{-1}\frac{\partial\hat K_1}{\partial\theta}+4\,e\,\hat a_0\frac{\partial\hat K_0}{\partial\theta}\nonumber\\
&&-8\,e^2L_2^2\,\hat c_0-\hat a_{-1}\,\frac{\partial\hat a_{-1}}{\partial\theta}\,\cot\theta+4\,e^2q_2\,\hat c_0+2\,\hat c_0\,\hat K_1\,\hat K_{-1}\nonumber\\
&&+2\,e\,\hat c_0\,\hat K_1-8\,m\,e\,\hat K_2-4\,e^2L_2\,\frac{\partial\hat K_1}{\partial\theta}+\hat a_0\,\hat c_{-1}\,\cot\theta\nonumber\\
&&+m\,\hat a_0\,\cot\theta-8\,m\,e\,L_2\,\hat a_0+4\,e\,L_2\,\hat a_{-1}\,\hat c_0+O_3\ .\label{122}
\end{eqnarray}
For simplicity we look for a model which is free from singularity for $0\leq\theta\leq\pi$ and involves only one arbitrary function of $u$, namely $E(u)$ describing the electric field experienced by the charged particle. Thus we can put the arbitrary functions $U(u), V(u)$ and $g(u)$ zero. When (\ref{122}) is written out explicitly we arrive at the following differential equation for $Q_2$:
\begin{eqnarray}
&&-\frac{1}{2}\underset{0}{\Delta}(\underset{0}{\Delta}Q_2+2\,Q_2)=\nonumber \\
&&\left (6\,m\,A+6\,e\,E-4\,e^2\dot A+14\,m\,e\,\dot E\right )\cos\theta \nonumber\\ 
&&+\biggl (2\,m^2E^2+6\,e^2A^2+30\,e^2E^2\biggr )(3\,\cos^2\theta-1)+O_3\ .\nonumber\\ \label{123}
\end{eqnarray}
The right hand side here is a linear combination of $l=1$ and $l=2$ Legendre polynomials. It can be integrated without encountering singularities at $\theta=0, \pi$ to provide us with the second order differential equation for $Q_2$:
\begin{eqnarray}
&&\underset{0}{\Delta}Q_2+2\,Q_2=(6\,m\,A+6\,e\,E-4\,e^2\dot A+14\,m\,e\,\dot E)\cos\theta\nonumber\\
&&+\left\{\frac{2}{3}\,m^2E^2+2\,e^2A^2+10\,e^2E^2\right\}(3\,\cos^2\theta-1)+O_3\ .\nonumber\\\label{125}
\end{eqnarray}
This will possess a particular integral which is singular at $\theta=0, \pi$ unless the coefficient of $\cos\theta$ (the $l=1$ Legendre polynomial) vanishes. Thus we must require that
\begin{equation}\label{125'}
m\,A=-e\,E+\frac{2}{3}\,e^2\dot A-\frac{7}{3}\,e\,m\,\dot E+O_3\ ,
\end{equation}
with the result that (\ref{125}) is solved by
\begin{equation}\label{126}
Q_2=-\frac{1}{2}\left\{\frac{1}{3}\,m^2E^2+e^2A^2+5\,e^2E^2\right\}(3\,\cos^2\theta-1)\ .
\end{equation}
We have not added a linear combination of $l=0$ and $l=1$ Legendre polynomials to this solution because such terms in $Q_2$ correspond to trivial perturbations as pointed out following (\ref{105}). 

In analyzing (\ref{125'}) we first note that the infinitesimal Lorentz transformation
\begin{equation}\label{127}
\bar v^3=v^3-\frac{7}{3}\,e\,E\,v^4+O_2\ ,\ \ \bar v^4=v^4-\frac{7}{3}\,e\,E\,v^3+O_2\ ,
\end{equation}
transforms away the $\dot E$--term in (\ref{125'}) since
\begin{equation}\label{128}
\bar A=A+\frac{7}{3}\,e\,\dot E+O_3\ .
\end{equation}
Then dropping the bars we have
\begin{equation}\label{129}
m\,A=-e\,E+\frac{2}{3}e^2\dot A+O_3\ .
\end{equation}
Next using $a^3=A\,v^4$ and $a^4=A\,v^3$ we have, on account of (\ref{106}),
\begin{eqnarray}
m\,a^3&=&e\,F^3{}_4\,v^4+\frac{2}{3}e^2\dot A\,v^4+O_3\ ,\label{130}\\
m\,a^4&=&e\,F^4{}_3\,v^3+\frac{2}{3}e^2\dot A\,v^3+O_3\ .\label{131}
\end{eqnarray}
But
\begin{eqnarray}
&&\dot A v^4=\dot a^3-A a^4=\dot a^3-A^2v^3=\dot a^3+(a^j a_j) v^3,\label{132}\\
&&\dot A v^3=\dot a^4-A a^3=\dot a^4-A^2v^4=\dot a^4+(a^j a_j) v^4,\label{132b}
\end{eqnarray}
and so, remembering that $v^1=0=v^2$, we can write (\ref{129}) in the equivalent form
\begin{equation}\label{133}
m\,a^i=e\,F^i{}_j\,v^j+\frac{2}{3}e^2\{\dot a^i+(a^j\,a_j)\,v^j\}+O_3\ .
\end{equation}
The first term on the right hand side here is the first order external 4--force (the Lorentz 4--force). The second term is the second order Lorentz--Dirac radiation reaction 4--force. There is no second order ``tail term" here because such a term is presumably inconsistent with maintaining rectilinear motion. We might have expected a second order external 4--force proportional to $e^2h^k_i\,F^p{}_k\,F_{pj}\,v^j$ where $h^k_i=\delta ^k_i-v^k\,v_i$ is the projection tensor (projecting 4--vectors orthogonal to $v^i$). However in the present case
\begin{eqnarray}
h^k_i\,F^p{}_k\,F_{pj}\,v^j&=&E^2(h^3_i\,v^3-h^4_i\,v^4)\nonumber\\
&=&E^2(\delta^3_i\,v^3-\delta^4_i\,v^4+v_i)=0\ .\label{135}
\end{eqnarray}

\section{Residual Matter Distribution}\label{sec_res_matter_dist}

Since the Einstein--Maxwell field equations have been satisfied approximately there exists a residual matter distribution described, in coordinates $x^{i'}=(\theta, \phi, r, u)$, by a 4--current $J^{i'}$ and an energy--momentum--stress tensor $T^{i'j'}$. We begin by examining the residual 4--current which is given by Maxwell's equations
\begin{equation}\label{136}
J^{i'}=F^{i'j'}{}_{;j'}\ .
\end{equation}
Thus in terms of the functions $F_1, F_2, F_3$ in (\ref{85})--(\ref{87}) the 4--current is given by
\begin{equation}\label{137}
\sqrt{-g}J^{i'}=\left (-\frac{\partial F_1}{\partial u}-\frac{\partial F_2}{\partial r},0,\frac{\partial F_2}{\partial\theta}+\frac{\partial F_3}{\partial u}, \frac{\partial F_1}{\partial\theta}-\frac{\partial F_3}{\partial r}\right ),
\end{equation}
with $\sqrt{-g}=r^2p^{-2}\sin\theta$. The evaluation of $F_1, F_2, F_3$ leading to (\ref{89})--(\ref{91}), and thus to the orders of magnitude (\ref{92})--(\ref{94}), can now be more explicit since we have found that $Q_1=0$ and we are in possession of the functions $\hat K_{-1}, \hat K_1, l_2, \hat a_{-1}$ and $\hat a_0$ in more explicit form. The result is
\begin{eqnarray}
&&F_1=\left (E\,\sin^2\theta-\frac{2}{3}e\,E^2\sin^2\theta\,\cos\theta+O_2\right )\,r\nonumber\\
&&+\biggl\{\frac{3}{2}\dot E\,\sin^2\theta+A\,E\,\sin^2\theta\,\cos\theta+O_1\biggr\}\,r^2\nonumber\\
&&+O(r^3)\ ,\label{138}\\
&&F_2=O_3\times\frac{1}{r}+\sin\theta\biggl\{\frac{\partial\hat K_0}{\partial\theta}+\left (2\,m\,E-e\,A-\frac{3}{2}e^2\dot E\right )\nonumber\\
&&\times\sin\theta+(8\,e^2A\,E-10\,e\,m\,E^2)\sin\theta\,\cos\theta+O_3\biggr\}\nonumber\\
&&+(3\,m\,\dot E\,\sin^2\theta-2\,e\,E^2\sin^2\theta\,\cos\theta+O_2)\,r\nonumber\\
&&+O(r^2)\ ,\label{139}\\
&&F_3=e\,\sin\theta-3\,e^2E\,\sin\theta\,\cos\theta+O_3\nonumber\\
&&+\left\{E\,\cos\theta-\frac{1}{3}e\,E^2(3\,\cos^2\theta-1)+O_2\right\}\,r^2\sin\theta\nonumber\\
&&+\biggl\{\dot E\,\cos\theta+\frac{1}{3}\,A\,E\,(3\,\cos^2\theta-1)+O_1\biggr\}\,r^3\sin\theta\nonumber\\
&&+O(r^4)\ ,\label{140}
\end{eqnarray}
where we have used (\ref{103}) to simplify the coefficient of $r$ in (\ref{139}). We can achieve the accuracy required in (\ref{966}) by replacing (\ref{95}) (with $U(u)=0$) by
\begin{eqnarray}
\frac{\partial\hat K_0}{\partial\theta}&=&(e\,A-2\,m\,E+3\,e^2\dot E)\sin\theta \nonumber \\
&+&(10\,e\,m\,E^2-8\,e^2A\,E)\sin\theta\,\cos\theta+O_3.\label{141x}
\end{eqnarray}
Thus the more accurate version of (\ref{96}) reads
\begin{eqnarray}
\hat K_0&=&-(e\,A-2\,m\,E+3\,e^2\dot E)\cos\theta\nonumber \\
&&+(5\,e\,m\,E^2-4\,e^2A\,E)\sin^2\theta+O_3\ ,\label{142x}
\end{eqnarray}
and (\ref{139}) is finally given by
\begin{eqnarray}
F_2&=&O_3\times\frac{1}{r}+\frac{3}{2}e^2\dot E\,\sin^2\theta+O_3+(3\,m\,\dot E\,\sin^2\theta\nonumber\\
&-&2\,e\,E^2\,\sin^2\theta\,\cos\theta+O_2)\,r+O(r^2)\ .\label{143x}
\end{eqnarray}
Now calculating $J^{i'}$ from (\ref{137}) we find that
\begin{eqnarray}
\sqrt{-g}\,J^{1'}&=&O_3\times\frac{1}{r^2}+2\,e\,E^2\sin^2\theta\,\cos\theta-3\,m\,\dot E\,\sin^2\theta\nonumber\\
&&+O_2+O(r)\ ,\label{144x}\\
\sqrt{-g}\,J^{2'}&=&0\ ,\label{145x}\\
\sqrt{-g}\,J^{3'}&=&O_3\times\frac{1}{r^2}+O_3\times\frac{1}{r}+\biggl\{6\,m\,\dot E\,\sin\theta\,\cos\theta\nonumber\\
&&-2\,e\,E^2\sin\theta\,(3\,\cos^2\theta-1)+O_2\biggr\}\,r+O(r^2)\ ,\nonumber \\ \label{146x}\\
\sqrt{-g}\,J^{4'}&=&O_2\times r+O_1\times r^2+O(r^3)\ ,\label{147x}
\end{eqnarray}
and from these the conservation equation for the 4--current takes the approximate form
\begin{equation}\label{148x}
\frac{\partial}{\partial x^{i'}}\left (\sqrt{-g}\,J^{i'}\right )=O_3\times\frac{1}{r^3}+O_3\times\frac{1}{r^2}+O_2+O(r)\ ,\end{equation}
which is a check on the trigonometric terms in (\ref{144x}) and (\ref{146x}). Solving (\ref{144x})--(\ref{147x}) for $J^{i'}$, using $p$ given by (\ref{79}) with $\hat P=1+O_2$, results in
\begin{eqnarray}
J^{1'}&=&O_3\times\frac{1}{r^4}+\{2\,e\,E^2\sin\theta\,\cos\theta-3\,m\,\dot E\,\sin\theta+O_2\}\nonumber\\
&&\times\frac{1}{r^2}+O\left (\frac{1}{r}\right )\ ,\label{141}\\
J^{2'}&=&0\ ,\label{142}\\
J^{3'}&=&O_3\times\frac{1}{r^3}+O_3\times\frac{1}{r^2}+\{6\,m\,\dot E\,\cos\theta\nonumber\\
&&-2\,e\,E^2(3\,\cos^2\theta-1)+O_2\}\times\frac{1}{r}+O(r^0)\ ,\label{143}\\
J^{4'}&=&O_2\times\frac{1}{r}+O_1+O(r)\ .\label{144}
\end{eqnarray}
We now introduce the half null tetrad defined via the 1--forms (\ref{31})--(\ref{34}) with $p, \alpha, a, c$ given by (\ref{79})--(\ref{82}). This consists of the covariant vectors (and their corresponding contravariant expressions):
\begin{eqnarray}
f_{i'}&=&(r\,p^{-1}e^{\alpha}, 0\ , 0\ , r\,p^{-1}a)\ \ \Leftrightarrow\ \ \nonumber\\
f^{i'}&=&(-r^{-1}p\,e^{-\alpha}, 0\ , 0\ , 0)\ ,\\
e_{i'}&=&(0, r\,p^{-1}e^{-\alpha}\sin\theta\ , 0\ , 0)\ \ \Leftrightarrow\ \ \nonumber\\
 e^{i'}&=&(0, -r^{-1}p\,e^{\alpha}\csc\theta\ , 0\ , 0)\ ,\\
l_{i'}&=&(0\ ,\ 0\ ,\ 1\ ,\ \frac{1}{2}\,c)\ \ \Leftrightarrow\ \ \nonumber\\
l^{i'}&=&(-a\,e^{-\alpha}\ ,\ 0\ ,\ -\frac{1}{2}\,c\ ,\ 1)\ ,\\
n_{i'}&=&(0\ ,\ 0\ ,\ 0\ ,\ 1)\ \ \Leftrightarrow\ \ n^{i'}=(0\ ,\ 0\ ,\ 1\ ,\ 0)\ .\nonumber\\\label{145}
\end{eqnarray}
The vectors $f^{i'}\ ,\ e^{i'}\ ,\ l^{i'}\ ,\ n^{i'}$ constitute a half null tetrad with $f^{i'}\ ,\ e^{i'}$ unit, orthogonal spacelike vectors and $l^{i'}\ ,\ n^{i'}$ two null vectors. All scalar products involving the four vectors vanish except $f_{i'}\,f^{i'}=e_{i'}\,e^{i'}=-l_{i'}\,n^{i'}=-1$. In terms of this basis we can write the 4--current $J^{i'}$ given by (\ref{141})--(\ref{144}) as
\begin{eqnarray}
J^{i'}&=&\Biggl\{O_3\times\frac{1}{r^3}+(3\,m\,\dot E\,\sin\theta-2\,e\,E^2\sin\theta\,\cos\theta \nonumber\\
&&+O_2)\times\frac{1}{r}+O(r^0)\Biggr\}\,f^{i'}+\Biggl\{O_3\times\frac{1}{r^3}+O_3\times\frac{1}{r^2} \nonumber\\
&&+\left (6\,m\,\dot E\,\cos\theta-2\,e\,E^2(3\,\cos^2\theta-1)+O_2\right )\times\frac{1}{r}\nonumber\\
&&+O(r^0)\Biggr\}\ n^{i'}+\Biggl\{O_2\times\frac{1}{r}+O_1+O(r)\Biggr\}\,l^{i'}\ .\label{146}\end{eqnarray}
To satisfy approximately the Einstein field equations (starting after (\ref{966}) above) we have worked with the tensor $W_{i'j'}=R_{i'j'}-2\,E_{i'j'}$. The residual energy--momentum--stress tensor $T_{i'j'}$ is given by Einstein's field equations:
\begin{equation}\label{147}
-8\,\pi\,T_{i'j'}=W_{i'j'}-\frac{1}{2}\,g_{i'j'}\,W=R_{i'j'}-\frac{1}{2}\,g_{i'j'}\,R-2\,E_{i'j'}\ .\end{equation}
Here $W=g^{i'j'}\,W_{i'j'}$, $R=g^{i'j'}\,R_{i'j'}$ is the Ricci scalar. The non--vanishing components $W_{i'j'}$ are found to be
\begin{eqnarray}
W_{1'1'}&=&O_3\times\frac{1}{r^2}+O_3+O_2\times r+O_1\times r^2+O(r^3)\ ,\nonumber\\\label{148}\\
W_{2'2'}&=&O_3\times\frac{1}{r^2}+O_3+O_2\times r+O_1\times r^2+O(r^3)\ ,\nonumber\\\label{149}\\
W_{3'3'}&=&O_1+O(r)\ ,\label{150}\\
W_{4'4'}&=&O_5\times\frac{1}{r^6}+O_4\times\frac{1}{r^5}+O_4\times\frac{1}{r^4}+O_3\times\frac{1}{r^3}\nonumber\\
&&+O_3\times\frac{1}{r^2}+\{6\,e\,\dot E\,\cos\theta+O_2\}\times\frac{1}{r}+O(r^0)\ ,\nonumber\\\label{151}\\
W_{1'3'}&=&O_3\times\frac{1}{r}+3\,e\,\dot E\,\sin\theta-2\,m\,A^2\sin\theta\,\cos\theta+O_2\nonumber\\
&&+O_1\times r+O(r^2)\ ,\label{152}\\
W_{1'4'}&=&O_3\times\frac{1}{r^3}+O_3\times\frac{1}{r^2}+\{-6\,e\,m\,\dot E\,\sin\theta\nonumber\\
&&+4\,m^2A^2\sin\theta\,\cos\theta+O_3\}\times\frac{1}{r}\nonumber\\
&&+\frac{3}{2}\,e\,\dot E\,\sin\theta+m\,A^2\,\sin\theta\,\cos\theta+O_2+O(r)\ ,\nonumber\\\label{153}\\
W_{3'4'}&=&O_3\times\frac{1}{r^4}+O_3\times\frac{1}{r^2}+\{3\,e\,\dot E\,\cos\theta\nonumber\\
&&-m\,A^2(3\,\cos^2\theta-1)+O_2\}\times\frac{1}{r}+O_1+O(r)\ .\nonumber\\\label{154}
\end{eqnarray}
Calculating $T_{i'j'}$ using (\ref{147}) and expressing the components in terms of the half null basis (\ref{145}) we arrive at
\begin{eqnarray}
8\,\pi\,T^{i'j'}&=&{\cal T}_1\,(f^{i'}\,n^{j'}+f^{j'}\,n^{i'})+{\cal T}_2\,n^{i'}\,n^{j'}+{\cal T}_3\,f^{i'}\,f^{j'}\nonumber\\
&&+{\cal T}_4\,e^{i'}\,e^{j'}+{\cal T}_5\,l^{i'}\,l^{j'}+{\cal T}_6\,(l^{i'}\,n^{j'}+l^{j'}\,n^{i'})\nonumber\\
&&+{\cal T}_7\,(f^{i'}\,l^{j'}+f^{j'}\,l^{i'})\ ,\label{155}
\end{eqnarray}
with ${\cal T}_1, $\dots$, {\cal T}_6$ given in Appendix \ref{app_formulas}.

To interpret the energy--momentum--stress tensor (\ref{155}) we consider it a tensor field on the background space--time in the neighborhood of the world line $r=0$ and we will neglect $O_3$--terms. To facilitate this we first express the basis vectors $f^{i'}, e^{i'}, l^{i'}, n^{i'}$ in terms of the vectors 
\begin{eqnarray}
\underset{(0)}{f}^{i'}&=&-r^{-1}\delta^{i'}_{1'}\ ,\ \underset{(0)}{e}^{i'}=-r^{-1}\csc\theta\,\delta^{i'}_{2'}\ ,\ \underset{(0)}{n}^{i'}=\delta^{i'}_{3'}\ ,\nonumber\\
\underset{(0)}{l}^{i'}&=&-A\,\sin\theta\,\delta^{i'}_{1'}+\left (-\frac{1}{2}+A\,r\,\cos\theta\right )\delta^{i'}_{3'}+\delta^{i'}_{4'}\ .\nonumber\\\label{163}
\end{eqnarray} 
These expressions are given exactly by 
\begin{eqnarray}
f^{i'}&=&p\,e^{-\alpha}\underset{(0)}{f}^{i'}\ ,\ e^{i'}=p\,e^{\alpha}\underset{(0)}{e}^{i'}\ ,\ n^{i'}=\underset{(0)}{n}^{i'}\ ,\nonumber\\
l^{i'}&=&\underset{(0)}{l}^{i'}+(a\,e^{-\alpha}-A\,\sin\theta)r\,\underset{(0)}{f}^{i'}\nonumber\\
&&+\left (-\frac{1}{2}c+\frac{1}{2}-r\,A\,\sin\theta\right )\underset{(0)}{n}^{i'}\ .\label{164}
\end{eqnarray}
When the expansions of $p, \alpha, a, c$ in powers of $r$ given by (\ref{79})--(\ref{82}) are substituted into (\ref{164}) and the results are in turn substituted into (\ref{155}) we arrive at (neglecting $O_3$--terms) the predominantly Vaidya form
\begin{eqnarray}
8\,\pi\,T^{i'j'}&=&\frac{1}{r^2}\{-6\,e\,m\,\dot E\,\cos\theta+2\,e^2E^2\sin\theta\,\cos\theta\}\underset{(0)}{n}^{i'} \underset{(0)}{n}^{j'}\nonumber\\
&&+O_1\times\frac{1}{r}+O(r^0)\ .\label{165}
\end{eqnarray}
Since we are working in the Minkowskian neighborhood of the world line $r=0$ we can write this in the rectangular Cartesian coordinates and time $X^i$, using $\underset{(0)}{n_{i'}}dx^{i'}=du=k_i\,dX^i$ which follows from (\ref{16}), as
\begin{eqnarray}
8\,\pi\,T^{ij}&=&\frac{1}{r^2}\{-6\,e\,m\,\dot E\,\cos\theta+2\,e^2E^2\sin\theta\,\cos\theta\}k^{i}\,k^{j}\nonumber\\
&&+O_1\times\frac{1}{r}+O(r^0)\ .\label{166}
\end{eqnarray}
The flux of 4--momentum across $r={\rm constant}$ in the direction of increasing $r$ and between the future null cones $u=u_1$ and $u=u_2$, with $u_2>u_1$ constants (see Fig.\ 1), is given by (\cite{Synge:1970} with our sign conventions)
\begin{equation}\label{167}
P^i=-r^2\int^{u_2}_{u_1}du\int T^{ij}r_{,j}\sin\theta\,d\theta\,d\phi\ ,
\end{equation}
with the integration with respect to $\theta, \phi$ over the ranges $0\leq\theta\leq\pi$ and $0\leq\phi\leq 2\pi$ respectively. With $k^i$ given by (\ref{10}) and the gradient of $r$ given by (\ref{16}) (remembering that $v^4=A\,v^3$ and $v^3=A\,v^4$) evaluation of (\ref{167}) using (\ref{166}) results in
\begin{equation}\label{168}
P^i=m^2\int_{u_1}^{u_2}\left (\frac{\dot A}{A}+\frac{\pi}{16}\,A\right )a^i\,du+O(r)=O_2+O(r)\ .
\end{equation}

Expressing the residual 4--current (\ref{146}) on the basis (\ref{163}), neglecting $O_3$--terms, and then changing from coordinates $x^{i'}$ to the rectangular Cartesians and time $X^i$, results in the residual 4--current in the Minkowskian neighborhood of $r=0$ in the background space--time being given by
\begin{eqnarray}
J^i&=&\frac{1}{r}\biggl\{\biggl (6\,m\,\dot E\,\cos\theta-3\,e\,E^2(3\,\cos^2\theta-1)\biggr )\,k^i\nonumber \\
&&+\biggl (-3\,m\,\dot E\,\sin\theta+2\,e\,E^2\sin\theta\,\cos\theta\biggr )\,\frac{\partial k^i}{\partial\theta}\biggr\}\nonumber\\
&&+O_2\times\frac{1}{r}+O(r^0)\ .\label{169}
\end{eqnarray}
Hence the total residual charge ${\cal C}$ crossing $r={\rm constant}$ in the direction of increasing $r$ and between the future null cones $u=u_1$ and $u=u_2$ is
\begin{equation}\label{170}
{\cal C}=-r^2\int_{u_1}^{u_2}\int J^i\,r_{,i}\,\sin\theta\,d\theta\,d\phi=O_2\times r+O(r^2)\ ,
\end{equation}
where we have used the gradient of $r$ given by (\ref{16}) and thus $k^i\,r_{,i}=1$ while $(\partial k^i/\partial\theta)\,r_{,i}=0$.

\section{Discussion}\label{sec_discussion}

It is interesting to compare the orders of magnitude of the fluxes of 4--momentum and charge (\ref{168}) and (\ref{170}) with the corresponding quantities in the case of the Bonnor--Vaidya particle. In this case the residual energy--momentum--stress tensor $T^{i'j'}$ and the residual 4--current $J^{i'}$ are given by (\ref{6}) and (\ref{7}). The background space--time is Minkowskian for $0\leq r<+\infty$ since there is no external field present. Considering (\ref{6}) and (\ref{7}) as tensor fields on the Minkowskian background and expressing them in terms of the rectangular Cartesians and time $X^i$, in the manner of section \ref{sec_res_matter_dist}, we have
\begin{equation}\label{171}
T^{ij}=\frac{1}{8\,\pi}\left (\frac{6\,m\,A\,\cos\theta}{r^2}-\frac{4\,e^2A\,\cos\theta}{r^3}\right )k^i\,k^j\ ,\end{equation}
and
\begin{equation}\label{172}
J^i=-\frac{2\,e\,A\,\cos\theta}{r^2}\,k^i\ .\end{equation}
In this case for $P^i$ and ${\cal C}$ we find, in place of (\ref{168}) and (\ref{170}),
\begin{equation}\label{173}
P^i=\left (m-\frac{2}{3}\frac{e^2}{r}\right )\int_{u_1}^{u_2}a^i\,du\ ,\end{equation}
and ${\cal C}=0$. If $e=0$ (Kinnersley case) we have ``the rocket effect" for which the total 4--momentum escaping across $r={\rm constant}$ in proper time $u_2-u_1$ is precisely the difference in the particle 4--momentum between the end and the beginning of this interval of proper time. This also applies to the Bonnor--Vaidya particle in the limit $r\rightarrow+\infty$ as can be seen from (\ref{173}). In our case however we can only compare (\ref{172}) and (\ref{173}) with (\ref{168}) and (\ref{170}) for small positive powers of $r$ and neglecting $O_3$--terms. We see that the introduction of an external field has removed ``the rocket effect" at the expense of no longer having arbitrary acceleration. Instead the acceleration is driven by the external field according to the important formula (\ref{133}). Helpful background to the approach adopted in this paper can be found in \cite{Hogan:Puetzfeld:2021}.

\begin{acknowledgments}
This work was funded by the Deutsche Forschungsgemeinschaft (DFG, German Research Foundation) through the grant PU 461/1-2 -- project number 369402949 (D.P.). 
\end{acknowledgments}

\appendix

\section{Useful formulas}\label{app_formulas}

For use in section \ref{sec_external_fields} the non--vanishing tetrad components of the Riemann tensor calculated on $r=0$ using the functions $q_2, c_2, a_1, \alpha_1$ are
\begin{eqnarray}
R_{(1)(2)(1)(2)}&=&\frac{1}{3}E^2(1-6\,\cos^2\theta)+C(3\,\cos^2\theta-1)\ ,\nonumber \\ \label{57}\\
R_{(1)(2)(2)(3)}&=&(2\,E^2-3\,C)\sin\theta\cos\theta\ ,\label{58}\\
R_{(1)(2)(2)(4)}&=&\left (-E^2+\frac{3}{2}C\right )\sin\theta\cos\theta\ ,\label{59}\\
R_{(1)(3)(1)(3)}&=&-3\,C\,\sin^2\theta\ ,\label{60}\\
R_{(1)(3)(1)(4)}&=&-\frac{1}{3}E^2+C\left (1-\frac{3}{2}\sin^2\theta\right )\ ,\label{61}\\
R_{(1)(3)(3)(4)}&=&3\,C\,\sin\theta\cos\theta\ ,\label{62}\\
R_{(1)(4)(1)(4)}&=&-\frac{3}{4}C\,\sin^2\theta\ ,\label{63}\\
R_{(1)(4)(3)(4)}&=&\frac{3}{2}C\,\sin\theta\cos\theta\ ,\label{64}\\
R_{(2)(3)(2)(3)}&=&(-2\,E^2+3\,C)\sin^2\theta\ ,\label{65}\\
R_{(2)(3)(2)(4)}&=&E^2\left (-\frac{1}{3}+\sin^2\theta\right )+C\left (1-\frac{3}{2}\sin^2\theta\right )\ ,\nonumber \\\label{66}\\
R_{(2)(4)(2)(4)}&=&\left (-\frac{1}{2}E^2+\frac{3}{4}C\right )\sin^2\theta\ ,\label{67}\\
R_{(3)(4)(3)(4)}&=&-\frac{1}{3}E^2-C(3\,\cos^2\theta-1)\ .\label{68}
\end{eqnarray}

For use in section \ref{sec_res_matter_dist} the components on the half null basis (\ref{145}) of the perturbed energy--momentum--stress tensor (\ref{155}) are
\begin{eqnarray}
{\cal T}_1&=&O_3\times\frac{1}{r^4}+O_3\times\frac{1}{r^3}+\{3\,e\,m\,\dot E\,\sin\theta -2\,e^2E^2\sin\theta\, \nonumber\\
&&\times \cos\theta+O_3\}\times\frac{1}{r^2}+\{2\,e\,A\,E\,\sin\theta\,\cos\theta+O_2\}\times\frac{1}{r} \nonumber\\
&&+O(r^0)\ ,\label{156}\\
{\cal T}_2&=&O_5\times\frac{1}{r^6}+O_4\times\frac{1}{r^5}+O_3\times\frac{1}{r^4}+O_3\times\frac{1}{r^3}\nonumber\\
&&+\{-6\,e\,m\,\dot E\,\cos\theta+2\,e^2E^2(3\,\cos^2\theta-1)+O_3\}\times\frac{1}{r^2}\nonumber\\
&&+\{-3\,e\,\dot E\,\cos\theta+e\,A\,E\,(3\,\cos^2\theta-1)+O_2\}\times\frac{1}{r}\nonumber\\
&&+O(r^0 )\ ,\label{157}\\
{\cal T}_3&=&O_3\times\frac{1}{r^4}+O_3\times\frac{1}{r^2}+\{-3\,e\,\dot E\,\cos\theta\nonumber\\
&&-e\,A\,E(3\,\cos^2\theta-1)+O_2\}\times\frac{1}{r}+O(r^0)\ ,\label{158}\\
{\cal T}_4&=&O_3\times\frac{1}{r^4}+O_3\times\frac{1}{r^2}+\{-3\,e\,\dot E\,\cos\theta\nonumber\\
&&-e\,A\,E(3\,\cos^2\theta-1)+O_2\}\times\frac{1}{r}+O(r^0)\ ,\label{159}\\
{\cal T}_5&=&O_1+O(r)\ ,\label{160}\\
{\cal T}_6&=&O_3\times\frac{1}{r^4}+O_3\times\frac{1}{r^2}+O_2\times\frac{1}{r}+O_1+O(r)\ ,\label{161}\\
{\cal T}_7&=&O_3\times\frac{1}{r^2}+\{-3\,e\,\dot E\,\sin\theta-2\,e\,A\,E\,\sin\theta\,\cos\theta\nonumber \\
&&+O_2\}\times\frac{1}{r}+O_1+O(r)\ .\label{162}
\end{eqnarray}

\bibliographystyle{unsrtnat}
\bibliography{accelcharge_bibliography}
\end{document}